\documentclass
[superscriptaddress,secnumarabic,amssymb,amsmath,nobibnotes,aps,prd,showpacs,showkeys,nofootinbib,onecolumn,12pt]{revtex4}%
\usepackage{graphicx}
\usepackage{subfigure}
\usepackage{epsf}
\usepackage{bm}
\usepackage{amsmath}
\usepackage{amsfonts}
\usepackage{amssymb}
\usepackage{color}%
\providecommand{\U}[1]{\protect\rule{.1in}{.1in}}

\newcommand{\be}{\begin{equation}}
\newcommand{\ee}{\end{equation}}

\newcommand{\mincir}{\raise
-3.truept\hbox{\rlap{\hbox{$\sim$}}\raise4.truept\hbox{$<$}\ }}
\newcommand{\magcir}{\raise
-3.truept\hbox{\rlap{\hbox{$\sim$}}\raise4.truept\hbox{$>$}\ }}

\begin{document}
\title{Modified Brans-Dicke cosmology with Minimum Length Uncertainty}
\author{Andronikos Paliathanasis}
\email{anpaliat@phys.uoa.gr}
\affiliation{Institute of Systems Science, Durban University of Technology, Durban 4000,
South Africa}
\affiliation{Instituto de Ciencias F\'{\i}sicas y Matem\'{a}ticas, Universidad Austral de
Chile, Valdivia, Chile}
\author{Genly Leon}
\email{genly.leon@ucn.cl}
\affiliation{Departamento de Matem\'{a}ticas, Universidad Cat\'{o}lica del Norte, Avda.
Angamos 0610, Casilla 1280 Antofagasta, Chile}

\begin{abstract}
We consider a modification of the Brans-Dicke gravitational Action Integral
inspired by the existence of a minimum length uncertainty for the scalar
field. In particular the kinetic part of the Brans-Dicke scalar field is
modified such that the equation of motion for the scalar field be modified
according \ to the quadratic Generalized Uncertainty Principle (GUP). For the
background geometry we assume the homogeneous and isotropic
Friedmann--Lema\^{\i}tre--Robertson--Walker metric. We investigate the
dynamics and the cosmological evolution for the dynamical variables of the
theory and we compare the results with the unmodified Brans-Dicke theory. It
follows that in consideration because of the additional degrees of freedom in
the energy-momentum tensor the dynamical variables describe various aspects of
the cosmological history. This is one of the first studies on the effects of
GUP in a Machian gravitational theory.

\end{abstract}
\keywords{Cosmology; Brans-Dicke; minimum length uncertainty; dynamical analysis}
\pacs{98.80.-k, 95.35.+d, 95.36.+x}
\date{\today}
\maketitle

\section{Introduction}

\label{sec1}

A systematic theoretical approach for the explanation of the recent
cosmological observations is the modification of the Einstein-Hilbert Action
Integral by the introduction of new degrees of freedom \cite{cl1}. Specifically, the newly
introduced dynamical terms in the Action Integral can describe matter
components, such as in the scalar field theories \cite{sf1,sf2,sf3,sf4}, or
geometrodynamical degrees of freedom with the introduction of geometric
invariants in the Action Integral \cite{mod1,mod2,mod3,mod4}. The purpose in
both approaches is common. The new dynamical terms in the field equations
should drive the dynamics \cite{mm1,mm2,mm3} such that to reproduce the
observational values for the physical parameters and describe various epochs
of the cosmological history \cite{ff1,ff2,ff3}.

Furthermore, the existence of a maximum energy scale in nature is predicted by
String Theory, the Doubly Special Relativity and other approaches of quantum
gravity \cite{ml1,ml2,ml3,ml4,ml5}. \ This specific energy scale follows as a
result of a minimum length scale, of the order of the Planck length, and the
modification of Heisenberg's Uncertainty Principle into a Generalized
Uncertainty principle (GUP) \cite{Maggiore}. The modification of the
Uncertainty Principle for the quantum observables leads to the definition of a
deformed Heisenberg algebra and consequently to the modification of the
Poisson Brackets on the classical limits. The effect of GUP in Hamiltonian
Mechanics and in General Relativity was the subject of study of a series of
works by various authors, see for instance \cite{mc1,mc2,mc3,mc4,mc5,mc6,mc7}
and references therein. There are various applications of GUP in cosmological
studies. Indeed, GUP has been applied as a mechanism for the description of
the cosmological constant as quantum effects \cite{cc1,cc2,cc3,cc4,cc5}. The
predicted value of the cosmological constant by the GUP \cite{cc5} fails to
explain the observations \cite{mc3}.

An alternate approach to the consideration of GUP in gravitational physics was
proposed in \cite{angup1}. In particular, it has been proposed that the
quintessence gravitational model \cite{sf1} for the description of dark
energy be modified such that the equation of motion for the scalar field,
i.e. the Klein-Gordon equation, include the higher-order derivative terms
provided by the deformed Heisenberg algebra. Indeed, in the case of the
quadratic GUP in the gravitational Action Integral of quintessence theory new
higher-order terms have been introduced as a result of the modification of the
Lagrangian function for the scalar field. In this approach the GUP has been
applied to modify the components in the gravitational Action Integral which
contributes to the energy-momentum tensor. With the use of a Lagrange
multiplier it has been found that the resulting theory is equivalent to that
of a multiscalar field model. The cosmological history and dynamics for the
modified quintessence model differ from the unmodified field where it was found
that the de Sitter Universe, that is, the exact solution with a cosmological
constant term, exists always in the dynamics independently of the scalar field
potential, which is not true for the unmodified model. Moreover, the effects
of the GUP\ can be observable in the cosmological perturbations, for more
details we refer the reader in \cite{angup2,angup3}.

In this piece of work inspired by \cite{angup1} we investigate the effects of
the GUP in the case of scalar-tensor theories and specifically in the case of
Brans-Dicke model \cite{Brans}. Such an analysis is important in order to
understand the effects of GUP under Mach's Principle. According to our
knowledge there are no applications in the literature on GUP of Mach's
Principle. Recall that why the initial attempt of Einstein is to construct a
Machian theory, General Relativity fails to satisfy Mach's Principle
\cite{ein1,ein2}, for instance Schwarzschild is vacuum solution which
describes a geometric object without any reference frame with inertia. The
plan of the paper is as follows.

In Section \ref{sec2} we present the basic equation for the Brans-Dicke
cosmology in the case of a spatially flat
Friedmann--Lema\^{\i}tre--Robertson--Walker (FLRW) background space. The
modified Brans-Dicke model is presented in Section \ref{sec3}. In Section
\ref{sec4} we present the dynamics and the cosmological evolution for the
physical variables as provided by the modified theory. The results are
compared with that of Brans-Dicke theory to understand the effects
of the GUP. Furthermore, in Section \ref{sec5} we extend our analysis in the
case of nonzero spatial curvature for the FLRW geometry. Finally, in Section
\ref{sec6} we summarize our results and we draw our conclusions.

\section{Brans-Dicke cosmology}

\label{sec2}

In 1961 \cite{Brans}, Carl H. Brans and Robert H. Dicke introduced a
gravitational theory which satisfies the Machian Principle. Specifically a
scalar field is introduced into the gravitational Action Integral which
interacts with the Ricci scalar. Indeed, for a four-dimensional Riemannian
space with metric term $g_{\mu\nu}$ and Ricci scalar $R$, the Brans-Dicke
Action Integral is expressed as
\begin{equation}
S=\int dx^{4}\sqrt{-g}\left[  \frac{1}{2}\phi R-\frac{1}{2}\frac{\omega_{BD}%
}{\phi}g^{\mu\nu}\phi_{;\mu}\phi_{;\nu}-V\left(  \phi\right)  \right],
\label{bd.01}%
\end{equation}
where $\omega_{BD}\neq \frac{3}{2}$ is known as the Brans-Dicke parameter,
into which a scalar field potential function, $V\left(  \phi\right)  $, has been introduced.

The gravitational field equations in Brans-Dicke Theory are%
\begin{equation}
G_{\mu\nu}=\frac{1}{\phi}\left(  \frac{\omega_{BD}}{\phi}\left(  \phi_{;\mu
}\phi_{;\nu}-\frac{1}{2}g_{\mu\nu}g^{\kappa\lambda}\phi_{;\kappa}%
\phi_{;\lambda}\right)  -g_{\mu\nu}V\left(  \phi\right)  -\left(  g_{\mu\nu
}g^{\kappa\lambda}\phi_{;\kappa\lambda}-\phi_{;\mu}\phi_{;\nu}\right)
\right), \label{bd.02}%
\end{equation}
in which the scalar field satisfies the equation of motion
\begin{equation}
g^{\mu\nu}\phi_{;\mu\nu}-\frac{1}{2\phi}g^{\mu\nu}\phi_{;\mu}\phi_{;\nu}%
+\frac{\phi}{2\omega_{BD}}\left(  R-2V_{,\phi}\right)  =0. \label{bd.03}%
\end{equation}

At this point it is important to mention that for large values of $\omega
_{BD}$, the contribution of the scalar field is small and one expects that, when
$\omega_{BD}$ reaches infinity, the limit of General Relativity is recovered.
However, as it was found \cite{omegaBDGR} Brans-Dicke theory and General
Relativity are different gravitational theories and the limit of General
Relativity is not recovered.

The Brans-Dicke Action Integral (\ref{bd.01}) is equivalent to another theory of
gravity. For $\omega_{BD}=0$, \ the Brans-Dicke Action reduces to the so-called
O'Hanlon theory of gravity \cite{Hanlon}. The latter is equivalent with the
higher-order theory of gravity known as $f\left(  R\right)  $ theory, where
the scalar field affects the higher-order derivatives by the use of a
Lagrange multiplier \cite{s01}.

Without loss of generality we can define the new field, $\ \phi=\psi^{2}$, such
that the Action Integral (\ref{bd.01}) is written as%
\begin{equation}
S=\int dx^{4}\sqrt{-g}\left[  \frac{1}{2}\psi^{2}R-\frac{\bar{\omega}_{BD}}%
{2}g^{\mu\nu}\psi_{;\mu}\psi_{;\nu}-V\left(  \psi\right)  \right],
\label{bd.04}%
\end{equation}
with $\omega_{BD}=4\bar{\omega}_{BD}$. Expression (\ref{bd.04}) belongs to the
family of scalar-tensor theories \cite{fbook}.

On the other hand, if we define the scalar field, $\phi=e^{\Phi}$, then the
Brans-Dicke Action (\ref{bd.01}) is%
\begin{equation}
S=\int dx^{4}\sqrt{-g}\left[  \frac{1}{2}e^{\Phi}\left(  R-\omega_{BD}%
g^{\mu\nu}\Phi_{;\mu}\Phi_{;\nu}\right)  -V\left(  \Phi\right)  \right],
\end{equation}
which is the Action for the dilaton field \cite{fbook}.

An important characteristic of the Scalar-tensor theories, consequently of the
Brans-Dicke theory, is that the Scalar-tensor theories are related though
conformal transformations, while the Scalar-tensor theories can be written in
the equivalent form of Einstein's General Relativity with a minimally coupled
scalar field under a conformal transformation. For more details we refer the
reader to \cite{ns11} and references therein.

\subsection{FLRW background space}

According to the cosmological principle, in large scales the background space
is described by the homogeneous and isotropic spatially flat FLRW metric with
metric tensor described by the line element%

\begin{equation}
ds^{2}=-dt^{2}+a^{2}\left(  t\right)  \left(  dr^{2}+r^{2}\left(  d\theta
^{2}+\sin^{2}\theta d\phi^{2}\right)  \right)  . \label{bd.05}%
\end{equation}

The function, $a\left(  t\right),  $ is the scale factor of the universe, while the
Hubble function is defined as $H=\frac{\dot{a}}{a},$ with~$\dot{a}=\frac
{da}{dt}$. Moreover, the scalar field is considered to inherit the symmetries
of the background space, that is, $\phi$ is homogeneous and depends only upon the
independent parameter $t$, that is, $\phi=\phi\left(  t\right)  $.

The gravitational field equations (\ref{bd.02}), (\ref{bd.03}) are
\cite{fbook}%

\begin{align}
3H^{2} &=\frac{\omega_{BD}}{2}\left(  \frac{\dot{\phi}}{\phi}\right)  ^{2}%
+\frac{V\left(  \phi\right)  }{\phi}-3H\left(  \frac{\dot{\phi}}{\phi}\right)
, \label{bd.001}%
\\
\dot{H}&=-\frac{\omega_{BD}}{2}\left(  \frac{\dot{\phi}}{\phi}\right)
^{2}+2H\left(  \frac{\dot{\phi}}{\phi}\right)  -\frac{1}{2(2\omega_{BD}%
+3)\phi}\left(  2V-\phi\frac{dV}{d\phi}\right), \label{bd.002}%
\\
\ddot{\phi}+3H\dot{\phi}& =\frac{1}{2\omega_{BD}+3}\left(  2V-\phi\frac
{dV}{d\phi}\right)  . \label{bd.003}%
\end{align}

The dynamical system (\ref{bd.001})-(\ref{bd.003}) has been widely studied in
the literature. Analytic and exact solutions have been found for various
functional forms of the potential function, $V\left(  \phi\right)  $, in a
series of studies \cite{sb01,sb02,sb03,sb04,sb05}. In the presence of
additional matter source exact and analytic solutions for the cosmological
field equations have been determined before in
\cite{sb06,sb06a,sb07,sb08,sb09}.

The dynamical evolution of the cosmological parameters and the construction of
the cosmological history provided by the field equations (\ref{bd.001}%
)-(\ref{bd.003}) was the subject of study in \cite{sb05,ref01}.  See also
\cite{ref3,ref4} for extensions.

For the quadratic potential function, $V\left(  \phi\right)  =V_{0}\phi^{2}$,
it was found \cite{ref01} that the field equations (\ref{bd.001}%
)-(\ref{bd.003}) provide two asymptotic solutions in which the scale factor is
power-law and a de Sitter solution which is always a future attractor. On the
other hand, when $V\left(  \phi\right)  =V_{0}\phi^{A}$, $A\neq2$, there exist
three asymptotic scaling solutions for the scale factor. For the analysis of
the dynamics for arbitrary potential see \cite{sb05}.

Another important characteristic for the dynamical system (\ref{bd.001}%
)-(\ref{bd.003}) \ is that it admits a minisuperspace description and it can be
reproduced by the variation of the point-like Lagrangian%
\begin{equation}
L\left(  a,\dot{a},\phi,\dot{\phi},\psi,\dot{\psi}\right)  =-6a\phi\dot{a}%
^{2}-6a^{2}\dot{a}\dot{\phi}-\frac{\omega_{BD}}{2\phi}a^{3}\dot{\phi}%
^{2}+a^{3}V\left(  \phi\right)  . \label{bd.004}%
\end{equation}

In the following we consider the generalization of the Brans-Dicke Action
Integral inspired by the GUP in which the minimum length uncertainty plays an
important role.

\section{Generalized Uncertainty Principle}

\label{sec3}

We consider the quadratic GUP where the modified Heisenberg uncertainty
principle is expressed as follows%
\begin{equation}
\Delta X_{i}\Delta P_{j}\geqslant\frac{\hbar}{2}[\delta_{ij}(1+\beta_{0}%
\frac{\ell_{Pl}^{2}}{2\hbar^{2}}P^{2})+2\beta_{0}\frac{\ell_{Pl}^{2}}%
{2\hbar^{2}}P_{i}P_{j}]\,,
\end{equation}
with deformed Heisenberg algebra
\begin{equation}
\lbrack X_{i},P_{j}]=i\hbar\lbrack\delta_{\alpha\beta}(1+\beta_{0}\frac
{\ell_{Pl}^{2}}{2\hbar^{2}}P^{2})+2\beta_{0}\frac{\ell_{Pl}^{2}}{2\hbar^{2}%
}P_{\alpha}P_{\beta}]. \label{pd.60}%
\end{equation}

The parameter, $\beta_{0}$, is called the deformation parameter
\cite{Quesne2006,Vagenas,Kemph1,Kemph2}, which can be positive or negative
\cite{neg1,neg2,neg3,neg5}. In the following we define $\beta=\beta_{0}%
\ell_{Pl}^{2}/2\hbar^{2}$. In the limit where $\beta_{0}=0$, the usual
Heisenberg uncertainty principle is recovered. The modification of the
Heisenberg uncertainty principle is not unique and other forms of the GUP have
been proposed in the literature \cite{sd1,sd2,sd3,sd4}.

In the relativistic four vector form, the commutation relation (\ref{pd.60})
can be written as \cite{Vagenas}
\begin{equation}
\lbrack X_{\mu},P_{\nu}]=-i\hbar\lbrack(1-\beta(\eta^{\mu\nu}P_{\mu}P_{\nu
}))\eta_{\mu\nu}-2\beta P_{\mu}P_{\nu}].\label{pd.61}%
\end{equation}
Consequently the deformed operators by keeping underformed the $X_{\mu}$ are
\begin{equation}
P_{\mu}=p_{\mu}(1-\beta(\eta^{\alpha\gamma}p_{\alpha}p_{\gamma}))~,~~X_{\nu
}=x_{\nu},\label{pd.62}%
\end{equation}
where now$~p^{\mu}=i\hbar\frac{\partial}{\partial x_{\mu}},$ and $[x_{\mu
},p_{\nu}]=-i\hbar\eta_{\mu\nu}$.

The Klein-Gordon equation for spin-0 particle with rest mass zero in the
concept of GUP is defined as%
\begin{equation}
\left[  \eta^{\mu\nu}P_{\mu}P_{\nu}-\left(  mc\right)  ^{2}\right]  \Psi=0,
\end{equation}
or equivalently by using (\ref{pd.62})
\begin{equation}
\square\Psi-2\beta\hbar^{2}\square\left(  \square\Psi\right)  +\left(
\frac{mc}{\hbar}\right)  ^{2}\Psi+O\left(  \beta^{2}\right)  =0.\label{pd.63}%
\end{equation}
$\square$ is the Laplace operator for the metric tensor $\eta_{\mu\nu}$.
Equation (\ref{pd.63}) is fourth-order partial differential equation.
Specifically it is a singular pertubation differential equation.

The Action Integral for the modified Klein-Gordon equation is
\begin{equation}
S_{KG}=\int dx^{4}\left(  \frac{1}{2}\eta^{\mu\nu}\mathcal{D}_{\mu}%
\Psi\mathcal{D}_{\nu}\Psi-\frac{1}{2}V_{0}\Psi^{2}\right),
\end{equation}
where $\mathcal{D}_{\mu}=\nabla_{\mu}+\beta\hbar^{2}\frac{\partial}{\partial
x^{\mu}}\square$.

Indeed we can introduce the new variable, $\Phi=\square\Psi$, to
write the modified Klein-Gordon equation as a system of two second-order
differential equations, that is%
\begin{equation}
\square\Psi-\Phi=0,
\end{equation}%
\begin{equation}
2\beta\hbar^{2}\square\Phi+\left(  V_{0}\Psi+\Phi\right)  =0,
\end{equation}
where now the resulting Lagrangian function is%
\begin{equation}
S_{KG}=\int dx^{4}\sqrt{-g}\left(  \frac{1}{2}\eta^{\mu\nu}\Psi_{,\mu}%
\Psi_{,\nu}+2\beta\hbar^{2}\eta^{\mu\nu}\Psi_{,\mu}\Phi_{,\nu}+\beta\hbar
^{2}\Phi^{2}-\frac{1}{2}V_{0}\Psi^{2}\right)  . \label{pd.64}%
\end{equation}

\subsection{Modified Brans-Dicke cosmology}

Inspired by the latter analysis, the modified scalar field Lagrangian function
is used \cite{gup1} as a dark energy candidate to modify the gravitational field
equations in the case of quintessence. Indeed, the field equations are of
higher-order.  Thus by the introduction of an additional scalar field the
modified Friedmann's equations are of second-order. It was found that because
of the presence of the perturbative terms the behaviour of the dynamics
differs and for the case of an exponential potential the de Sitter universe
follows, in contrast to the usual quintessence scenario \cite{gup2}.  This
means that quadratic GUP may play role in the description of the inflationary
era. The relation between the cosmological constant and the GUP has been
investigated before \cite{sd2,cc01,cc02}; however our approach is different.
In our approach we introduce new degrees of freedom by modify the Lagrangian
for the matter component

Without loss of generality we assume the Brans-Dicke Action Integral
(\ref{bd.04}).  Then by following \cite{gup1} with the use of (\ref{pd.64}) it
follows that%
\begin{equation}
S_{mBD}=\int dx^{4}\sqrt{-g}\left[  \frac{1}{2}\psi^{2}R-\frac{\bar{\omega
}_{BD}}{2}\left(  g^{\mu\nu}\nabla_{\mu}\psi\nabla_{\nu}\psi+2\beta\hbar
^{2}\left(  2g^{\mu\nu}\nabla_{\mu}\psi\nabla_{\nu}\zeta+\zeta^{2}\right)
\right)  +V\left(  \psi\right)  \right],\label{pd.65}%
\end{equation}
where $\zeta=\Delta\psi$, $\Delta$ is the Laplace operator with respect to the
metric tensor $g_{\mu\nu}$.

Hence, for the FLRW background space with line element (\ref{bd.05}) from the
Action Integral (\ref{pd.65}) we derive the modified Brans-Dicke point-like
Lagrangian%
\begin{equation}
L\left(  a,\dot{a},\phi,\dot{\phi},\psi,\dot{\psi}\right)  =-6a\psi^{2}\dot
{a}^{2}-12\psi a^{2}\dot{a}\dot{\psi}-\frac{\bar{\omega}_{BD}}{2}a^{3}\left(
\dot{\psi}^{2}+2\beta\hbar^{2}\dot{\psi}\dot{\zeta}-\zeta^{2}\right)
+a^{3}V\left(  \psi\right)  .\label{pd.66}%
\end{equation}

Therefore, the modified Brans-Dicke field equations are%
\begin{equation}
6H^{2}+12H\left(  \frac{\dot{\psi}}{\psi}\right)  +\frac{\bar{\omega}_{BD}}%
{2}\left(  \dot{\psi}^{2}+2\beta\hbar^{2}\left(  \frac{\dot{\psi}}{\psi
}\right)  \left(  \frac{\dot{\zeta}}{\psi}\right)  -\frac{\zeta^{2}}{\psi^{2}%
}\right)  +\frac{V\left(  \psi\right)  }{\psi^{2}}=0, \label{pd.01}%
\end{equation}%
\begin{equation}
2\dot{H}+3H^{2}+4\left(  \frac{\dot{\psi}}{\psi}\right)  H+\left(  \frac
{\bar{\omega}_{BD}}{4}-2\right)  \left(  \frac{\dot{\psi}}{\psi}\right)
^{2}+2\left(  \frac{\ddot{\psi}}{\psi}\right)  +\frac{V\left(  \psi\right)
}{2\psi^{2}}+\beta\frac{\bar{\omega}_{BD}}{4\psi^{2}}\left(  \zeta^{2}%
-2\dot{\zeta}\dot{\psi}\right)  =0, \label{pd.02}%
\end{equation}%
\begin{equation}
\ddot{\psi}+3H\dot{\psi}+\zeta=0 \label{pd.03}%
\end{equation}
and%
\begin{equation}
\bar{\omega}_{BD}\left(  \beta\ddot{\zeta}+\ddot{\psi}\right)  +3\bar{\omega
}_{BD}H\left(  \beta\dot{\zeta}+\dot{\psi}\right)  +V_{,\psi}+12\psi\left(
\dot{H}+2H^{2}\right)  =0. \label{pd.04}%
\end{equation}

We continue our analysis by studying the cosmological evolution and dynamics for
dynamical variables as provided by the modified Brans-Dicke cosmological field equations.

\section{Dynamical analysis}

\label{sec4}

In order to study the dynamics of the field equations we work in the
$H$-normalization. Thus we define the new set of variables \cite{ref4}
\begin{equation}
x=\frac{\dot{\psi}}{\psi H}~,~y=\frac{V\left(  \psi\right)  }{6\psi^{2}H^{2}%
}~,~z=\beta\hbar^{2}\frac{\dot{\zeta}}{6\psi^{2}H^{2}}~,~w=\beta\hbar^{2}%
\frac{\zeta^{2}}{12\psi^{2}H}~, \label{pd.05}%
\end{equation}%
\begin{equation}
\lambda=\psi\frac{V_{,\psi}\left(  \psi\right)  }{V\left(  \psi\right)
}~,~\mu=-\frac{1}{\beta}\frac{\psi}{\zeta}~,~\Gamma\left(  \lambda\right)
=\frac{V_{,\psi\psi}\left(  \psi\right)  V\left(  \psi\right)  }{\left(
V_{,\psi}\left(  \psi\right)  \right)  ^{2}}.  \label{pd.06}%
\end{equation}

Hence, the field equations, (\ref{pd.01})-(\ref{pd.04}), can be written as the
following algebraic-differential system%
\begin{align}
\frac{dx}{d\tau}  &  =\left(  1-\frac{\bar{\omega}_{BD}}{8}\right)
x^{3}-\frac{1}{2}\left(  4+3\bar{\omega}_{BD}z\right)  x^{2}\nonumber\\
&  +6\mu w-\frac{3}{2}x\left(  1-y-\left(  \bar{\omega}_{BD}+4\mu\right)
w\right), \label{pd.07}%
\\
\frac{4}{y}\frac{dy}{d\tau}  &  =\left(  1-\bar{\omega}_{BD}\right)
x^{2}+12\left(  1+y\right) \nonumber\\
&  +4x\left(  \lambda-4-3\bar{\omega}_{BD}z\right)  +12\left(  \bar{\omega
}_{BD}+4\mu\right)  w~, \label{pd.08}%
\end{align}
\begin{align}
8\bar{\omega}_{BD}\frac{dz}{d\tau}  &  =8\left(  1+\left(  \lambda-3\right)
y\right)  +12\bar{\omega}_{BD}\left(  1-y\right)  z\nonumber\\
&  +\left(  \bar{\omega}_{BD}-8\right)  \left(  2+\bar{\omega}_{BD}z\right)
x^{2}-3\bar{\omega}_{BD}\left(  \bar{\omega}_{BD}+4\mu\right)  z\nonumber\\
&  +4x\left(  4+\bar{\omega}_{BD}\left(  10+3\bar{\omega}_{BD}z\right)
z\right)  +4w\left(  2\left(  \bar{\omega}_{BD}-12\right)  \mu-6\bar{\omega
}_{BD}\right), \label{pd.09}%
\\
\frac{4}{w}\frac{dw}{d\tau}  &  =12\left(  1+y\right)  -x\left(  16+\left(
\bar{\omega}_{BD}-8\right)  x+12\bar{\omega}_{BD}z\right) \nonumber\\
&  -12\left(  2\mu z-\left(  \bar{\omega}_{BD}+4\mu\right)  w\right),
\label{pd.10}%
\\
\frac{d\lambda}{d\tau} &=\lambda x\left(  1-\lambda\left(  1-\Gamma\left(
\lambda\right)  \right)  \right), \label{pd.11}%
\\
\frac{d\mu}{d\tau}& =\mu\left(  x+3z\mu\right), \label{pd.12}%
\end{align}%
with algebraic constraint%
\begin{equation}
1-\omega_{BD}w+2x+\frac{1}{12}\omega_{BD}x^{2}+y+\omega_{BD}xz=0. 
\label{pd.14}%
\end{equation}

Furthermore, the equation of state parameter for the effective cosmological
fluid is expressed as%
\begin{equation}
w_{tot}\left(  x,y,z,w,\lambda,\mu\right)  =y-\frac{1}{12}x\left(  8+x\left(
\bar{\omega}_{BD}-8\right)  +12\bar{\omega}_{BD}z\right)  +w\left(
\bar{\omega}_{BD}+4\mu\right). \label{pd.15}%
\end{equation}

With the use of the algebraic equation (\ref{pd.14}) the dynamical system
(\ref{pd.07})-(\ref{pd.12}) can be reduced from a six-dimensional system to a
five-dimensional system. Moreover, for the scalar field potential $V\left(
\psi\right)  $ we consider the power-law potential function $V\left(
\psi\right)  =V_{0}\psi^{\lambda_{0}}$, such that equation (\ref{pd.11})
become $\frac{d\lambda}{d\tau}=0$, with $\lambda=\lambda_{0}$. In such a case
the dynamical system of our consideration is reduced to a four-dimensional system.

Let $P=\left(  x\left(  P\right),y\left(  P\right),z\left(  P\right)
,\mu\left(  P\right)  \right)  $ be a stationary point for the system
composed of the equations (\ref{pd.07}), (\ref{pd.08}), (\ref{pd.09}) and
(\ref{pd.12}). Then by definition at the stationary point, $P$, the rhs of
equations (\ref{pd.07}), (\ref{pd.08}), (\ref{pd.09}) and (\ref{pd.12}) are zero.

The stationary points are
\[
P_{1}=\left(  -1,0,z_{1},0\right),~P_{2}=\left(  -\frac{6}{\lambda}%
,\frac{2}{\lambda^{2}}\left(  \lambda-6\right),z_{2},0\right),
\]%
\[
P_{3}=\left(  0,-1,\frac{\lambda-4}{3\bar{\omega}_{BD}},0\right)
~,~P_{4}=\left(  -1,0,\frac{1}{12}-\frac{1}{\bar{\omega}_{BD}},\frac
{4\bar{\omega}_{BD}}{\bar{\omega}_{BD}-4}\right),
\]%
\[
P_{5}=\left(  -\frac{6}{\lambda},\frac{2}{\lambda^{2}}\left(  \lambda
-6\right),\frac{\lambda\left(  \lambda-10\right)  +3\left(  \bar{\omega
}_{BD}-4\right)  }{6\bar{\omega}_{BD}\lambda},\frac{12\bar{\omega}_{BD}%
}{\lambda\left(  \lambda-10\right)  +3\left(  \bar{\omega}_{BD}-4\right)
}\right),
\]%
\[
P_{6}=\left(  x_{6},0,-\frac{2}{\bar{\omega}_{BD}}\left(  2+\frac{1}{x_{6}%
}\right)  -\frac{x_{6}}{6},\frac{2\bar{\omega}_{BD}x_{6}^{2}}{12+x_{6}\left(
24+\bar{\omega}_{BD}x_{6}\right)  }\right).
\]

Points $P_{1}$ describe describe a family of radiation-like solutions defined
on the the three dimensional space with arbitrary value of $z$. The
equation of state parameter is $w_{tot}\left(  P_{1}\right)  =\frac{1}{3}$,
and the scale factor of the asymptotic solution is expressed by the power-law
function $a\left(  t\right)  =a_{0}t^{\frac{1}{2}}$. \ In order to infer
the stability properties of the point we determine the eigenvalues of the
linearized system around $P_{1}$. The four eigenvalues are%
\[
e_{1}\left(  P_{1}\right)  =0\text{~},~e_{2}\left(  P_{1}\right)
=1,~e_{3}\left(  P_{1}\right)  =-1~,~e_{4}\left(  P_{1}\right)  =6-\lambda~,
\]
which means that the points $P_{1}$ are always a saddle points.

The family of points $P_{2}$ describes scaling solutions with $w_{tot}\left(
P_{2}\right)  =1-\frac{4}{\lambda}$. The asymptotic solution describes an
accelerated universe for $0<\lambda<3$. The eigenvalues of the linearized
system are%
\[
e_{1}\left(  P_{2}\right)  =0~,~e_{2}\left(  P_{2}\right)  =-\frac{6}{\lambda
}~,
\]%
\[
~e_{3}\left(  P_{2}\right)  =\frac{3}{\lambda}\left(  1+\sqrt{25-4\lambda
}\right),~e_{4}\left(  P_{2}\right)  =\frac{3}{\lambda}\left(
1-\sqrt{25-4\lambda}\right)  .
\]
Hence, points $P_{2}$ are always saddle points.

The asymptotic solution at the point $P_{3}$ is that of the de Sitter
universe, i.e. $w_{tot}\left(  P_{3}\right)  =-1$ and $a\left(  t\right)
=a_{0}e^{H_{0}t\text{.}}$ We remark that the de Sitter solutions exist for
arbitrary values of the parameter, $\lambda$, in contrast to the classical
Brans-Dicke theory for which solutions exist only of a specific value of $\lambda$. That
result is similar of the modified from the GUP quintessence model
\cite{gup1}. The eigenvalues of the linearized system are derived to be
\[
e_{1}\left(  P_{3}\right)  =0~,~e_{2}\left(  P_{3}\right)  =-6~,
\]%
\[
e_{3}\left(  P_{3}\right)  =-3\text{~},~e_{4}\left(  P_{3}\right)  =-3. 
\]
Thus, there exists a codimension one surface on which the de Sitter universe is an attractor (the stable manifold). In
order to determine the stability in the full phase space the center manifold theorem is applied because one eigenvalue is zero (then, exists a one dimensional center manifold).

Point $P_{4}$ corresponds to the radiation solution with $w_{tot}\left(
P_{4}\right)  =\frac{1}{3}$. The eigenvalues are
\[
e_{1}\left(  P_{4}\right)  =1~,~e_{2}\left(  P_{4}\right)  =1~,
\]%
\[
e_{3}\left(  P_{4}\right)  =2~,~e_{4}\left(  P_{4}\right)  =6-\lambda
\text{~},
\]
from which we infer that the stationary point $P_{4}~$is a source for
$\lambda<6$ or a saddle point for $\lambda>6$.

Point $P_{5}$ corresponds to the scaling solutions with $w_{tot}\left(
P_{5}\right)  =1-\frac{4}{\lambda}$. The eigenvalues of the linearized system
around the stationary point are
\[
e_{1}\left(  P_{5}\right)  =\frac{6}{\lambda}~\text{,~}e_{2}\left(
P_{5}\right)  =\frac{12}{\lambda}%
\]%
\[
~e_{3}\left(  P_{2}\right)  =\frac{3}{\lambda}\left(  1+\sqrt{25-4\lambda
}\right),~e_{4}\left(  P_{2}\right)  =\frac{3}{\lambda}\left(
1-\sqrt{25-4\lambda}\right)
\]
from which we infer that $P_{5}$ is always a saddle point.

Finally, point $P_{6}$ describes a family of points for which the asymptotic
solution has $w_{tot}\left(  P_{6}\right)  =1+\frac{4}{3}x_{6}$. Hence, for
$x_{6}<-1$, the solution describes an accelerated universe. The eigenvalues
are determined to be%
\[
e_{1}\left(  P_{6}\right)  =0~,~e_{2}\left(  P_{6}\right)  =6+x_{6}\left(
2+\lambda\right),
\]%
\[
e_{3}\left(  P_{6}\right)  =x_{6}\sqrt{\frac{\left(  \bar{\omega}%
_{BD}-36\right)  \left(  x_{6}\right)  ^{2}-24\left(  1+2x_{6}\right)
}{12+24x_{6}+\bar{\omega}_{BD}\left(  x_{6}\right)  ^{2}}}~,
\]%
\[
e_{5}\left(  P_{6}\right)  =-x_{6}\sqrt{\frac{\left(  \bar{\omega}%
_{BD}-36\right)  \left(  x_{6}\right)  ^{2}-24\left(  1+2x_{6}\right)
}{12+24x_{6}+\bar{\omega}_{BD}\left(  x_{6}\right)  ^{2}}.}%
\]
Therefore point $P_{6}$ is a saddle point. The results are summarized in Table
\ref{tab1}.%

\begin{table}[tbp] \centering
\caption{Stationary points for the modified Brans-Dicke cosmological model.}%
\begin{tabular}
[c]{cccc}\hline\hline
\textbf{Point} & $\mathbf{w}_{tot}\left(  \mathbf{P}\right)  $ &
\textbf{Acceleration} & \textbf{Stability}\\\hline
$P_{1}$ & $\frac{1}{3}$ & No & Saddle\\
$P_{2}$ & $1-\frac{4}{\lambda}$ & $0<\lambda<3$ & Saddle\\
$P_{3}$ & $-1$ & Yes & Saddle\\
$P_{4}$ & $\frac{1}{3}$ & No & Saddle\\
$P_{5}$ & $1-\frac{4}{\lambda}$ & $0<\lambda<3$ & Saddle\\
$P_{6}$ & $1+\frac{4}{3}x_{6}$ & $x_{6}<-1$ & Saddle\\\hline\hline
\end{tabular}
\label{tab1}%
\end{table}%

Let us now compare these results with those of the standard Brans-Dicke theory
for the power-law potential. In standard Brans-Dicke theory, there exist only
three stationary points which describe scaling solutions.\ The equation of
state parameters at these points depend upon the power $\lambda$ of the
potential and on the Brans-Dicke parameter $\omega_{BD}$. The de Sitter
solution exists only for a specific parameter of $\lambda$. On the other hand
in the model proposed in this work the field equations admit six stationary
points. Three of the points, $P_{1}$, $P_{2}$ and $P_{6}$ are actually
families of points. The asymptotic solutions describe three scaling solutions,
two radiation solutions and the de Sitter universe. It is important to mention
that the corresponding cosmological fluid at the scaling solutions, i.e. the
parameters for the equation of state, does not depend upon the Brans-Dicke
parameter. Specifically $w_{tot}$ is a function of the parameter $\lambda$ and
of the fixed coordinates on the phase space of the stationary point through equation \eqref{pd.15}.

\subsection{Center manifold analysis for point $P_{3}$}

\subsubsection{Case $\lambda \neq 4$}

The Jordan Decomposition of the Jacobian matrix evaluated at $P_{3}$,
\begin{equation}
 m=   \left(
\begin{array}{cccc}
 -3 & 0 & 0 & 0 \\
 -\lambda -2 & -6 & 0 & 0 \\
 \frac{\lambda +8}{3 \bar{\omega}_{BD}} & \frac{2}{\bar{\omega}_{BD}} & -3 & 0 \\
 0 & 0 & 0 & 0 \\
\end{array}
\right),
\end{equation}
We assume that $\lambda \neq 4$.  The matrix $m$
can be decomposed as
\begin{equation}
 m = s \cdot j \cdot s^{-1},
\end{equation}
with similarity matrix
\begin{equation}
 s=    \left(
\begin{array}{cccc}
 0 & 0 & -\frac{3 \bar{\omega}_{BD}}{\lambda -4} & 0 \\
 -\frac{3 \bar{\omega}_{BD}}{2} & 0 & \frac{(\lambda +2) \bar{\omega}_{BD}}{\lambda -4} & 0 \\
 1 & 1 & 0 & 0 \\
 0 & 0 & 0 & 1 \\
\end{array}
\right),
\end{equation}
and $j$ is the Jordan canonical form of $m$:
\begin{equation}
    j= \left(
\begin{array}{cccc}
 -6 & 0 & 0 & 0 \\
 0 & -3 & 1 & 0 \\
 0 & 0 & -3 & 0 \\
 0 & 0 & 0 & 0 \\
\end{array}
\right).
\end{equation}
We introduce the linear transformation
\begin{equation}
 \left(\begin{array}{c}
      v   \\
      y_1 \\
      y_2 \\
      y_3
 \end{array}\right)  =  \left(\begin{array}{c}
 \mu\\
 -\frac{2 ((\lambda +2) x+3 (y+1))}{9 \bar{\omega}_{BD}}\\
    \frac{-3 \lambda +2 (\lambda +2) x+6 y+9 \bar{\omega}_{BD} z+18}{9 \bar{\omega}_{BD}}\\
    -\frac{(\lambda -4)
   x}{3 \bar{\omega}_{BD}}\end{array}
\right). \label{linear}
\end{equation}
Then,  assuming $\lambda\neq 4$, and using the linear transformation \eqref{linear} we can write the dynamical system
(\ref{pd.07})-(\ref{pd.12}) as
\begin{align}
& \frac{d v}{d\tau}= 3 y_{1} v^2+3 y_{2} v^2-\frac{3 y_{3} v \bar{\omega}_{BD}}{\lambda -4}+\frac{(\lambda -4) v^2}{\bar{\omega}_{BD}},  \label{eq.46}
\\
 & \frac{d y_1}{d\tau}= -6 y_1 +  y_{1}^2 \left(-\frac{36 y_{3} v \bar{\omega}_{BD}}{\lambda -4}-18 v-9 \bar{\omega}_{BD}\right) \nonumber \\
 & +y_{1} \Bigg(-\frac{36 y_{2} y_{3} v
   \bar{\omega}_{BD}}{\lambda -4}+y_{3}^2 \left(\frac{18 \bar{\omega}_{BD}^2}{(\lambda -4)^2}+v \left(\frac{9 \bar{\omega}_{BD}^2}{(\lambda -4)^2}+\frac{12 (\lambda +2)
   \bar{\omega}_{BD}}{(\lambda -4)^2}\right)\right) \nonumber \\
   & +y_{3} \left(\frac{6 (\lambda +2) \bar{\omega}_{BD}}{\lambda -4}+\left(\frac{36}{\lambda -4}+10\right) v\right)+\frac{2
   (\lambda -4) v}{\bar{\omega}_{BD}}\Bigg)  +y_{2} \left(\frac{12 (\lambda +2) y_{3}^2 v \bar{\omega}_{BD}}{(\lambda -4)^2}+4 y_{3}
   v\right) \nonumber \\
   & +y_{3}^3 \left(-\frac{6 (\lambda +2) \bar{\omega}_{BD}^2}{(\lambda -4)^3}-\frac{3 (\lambda +2) v \bar{\omega}_{BD}^2}{(\lambda -4)^3}\right)+y_{3}^2
   \left(\frac{v \bar{\omega}_{BD}}{4-\lambda }-\frac{2 \bar{\omega}_{BD}}{\lambda -4}\right), \label{eq.43}
\\
    & \frac{d y_2}{d\tau}=-3 y_2 + y_{1}^2 \left(\frac{18 y_{3} v \bar{\omega}_{BD}}{\lambda -4}+9
   v+\frac{9 \bar{\omega}_{BD}}{2}\right) \nonumber \\
   & +y_{1} \Bigg(y_{2} \left(-9 v-\frac{9 \bar{\omega}_{BD}}{2}\right)+y_{3}^2 \left(v \left(-\frac{9 \bar{\omega}_{BD}^2}{2 (\lambda -4)^2}-\frac{12 (\lambda +2) \bar{\omega}_{BD}}{(\lambda -4)^2}\right)-\frac{9 \bar{\omega}_{BD}^2}{(\lambda -4)^2}\right)\nonumber \\
   & +y_{3} \left(v \left(\frac{3
   \bar{\omega}_{BD}}{\lambda -4}-\frac{8 (2 \lambda +1)}{\lambda -4}\right)-\frac{3 (\lambda +3) \bar{\omega}_{BD}}{\lambda -4}\right)+v \left(\frac{2-5 \lambda }{\bar{\omega}_{BD}}+\frac{3}{2}\right)\Bigg)  -\frac{18 y_{2}^2 y_{3} v \bar{\omega}_{BD}}{\lambda -4} \nonumber \\
   & +y_{2} \Bigg(y_{3}^2 \left(\frac{9 \bar{\omega}_{BD}^2}{(\lambda
   -4)^2}+v \left(\frac{9 \bar{\omega}_{BD}^2}{2 (\lambda -4)^2}-\frac{12 (\lambda +2) \bar{\omega}_{BD}}{(\lambda -4)^2}\right)\right) \nonumber \\
   & +y_{3} \left(\frac{3 (\lambda +1)
   \bar{\omega}_{BD}}{\lambda -4}+v \left(\frac{3 \bar{\omega}_{BD}}{\lambda -4}+\frac{4-10 \lambda }{\lambda -4}\right)\right)\Bigg)  +y_{3}^3 \left(\frac{6 (\lambda +2)
   \bar{\omega}_{BD}^2}{(\lambda -4)^3}+\frac{3 (\lambda +2) v \bar{\omega}_{BD}^2}{(\lambda -4)^3}\right) \nonumber \\
   & +y_{3}^2 \left(\frac{(5 \lambda -2) \bar{\omega}_{BD}}{(\lambda
   -4)^2}+v \left(\frac{(5 \lambda -2) \bar{\omega}_{BD}}{2 (\lambda -4)^2}-\frac{3 \bar{\omega}_{BD}^2}{4 (\lambda -4)^2}\right)\right)+y_{3},  \label{eq.44}
\\
     & \frac{d y_3}{d\tau}=-3 y_{3} + y_{1} \left(-\frac{18
   y_{3}^2 v \bar{\omega}_{BD}}{\lambda -4}+y_{3} \left(-3 v-\frac{9 \bar{\omega}_{BD}}{2}\right)+\frac{3 (\lambda -4) v}{\bar{\omega}_{BD}}\right) \nonumber \\
   & +y_{2} \left(6 y_{3} v-\frac{18 y_{3}^2 v \bar{\omega}_{BD}}{\lambda -4}\right)+y_{3}^3 \left(\frac{9 \bar{\omega}_{BD}^2}{(\lambda
   -4)^2}+\frac{9 v \bar{\omega}_{BD}^2}{2 (\lambda -4)^2}\right)   +y_{3}^2 \left(\frac{3 (\lambda +1) \bar{\omega}_{BD}}{\lambda -4}+\frac{3 v \bar{\omega}_{BD}}{8-2
   \lambda }\right). \label{eq.45}
\end{align}
The system \eqref{eq.46}, \eqref{eq.43}, \eqref{eq.44}, \eqref{eq.45},  is written in matrix form as
\begin{align}
 & \frac{d v}{d \tau}= C v + f(x, \mathbf{y}), \label{eq.47}\\
 &   \frac{d\mathbf{y}}{d \tau}= P \mathbf{y} + \mathbf{g}(v, \mathbf{y}), \quad \mathbf{y}=(y_1,y_2, y_3)^T, \label{eq.48}
\end{align}
where
\begin{align}
C=0, \quad  P=   \left(
\begin{array}{ccc}
 -6 & 0 & 0  \\
 0 & -3 & 1  \\
 0 & 0 & -3
\end{array}
\right).
\end{align}
That is, $C$ is the zero $1\times 1$ matrix, $P$ is the square matrix with negative eigenvalues, $f$ and  $ \mathbf{g}$ vanishes at $\mathbf{0}$ and have vanishing derivatives at $\mathbf{0}$. The center manifold theorem asserts that there is a 1-dimensional invariant local center manifold $W^c(\mathbf{0})$ of  \eqref{eq.47}-\eqref{eq.48} tangent to the center subspace (the $\mathbf{y}=\mathbf{0}$ space) at  $\mathbf{0}$. Moreover, $W^c(\mathbf{0})$ can be represented as
\begin{equation}
    W^c(\mathbf{0})=\left\{(v, \mathbf{y})\in \mathbb{R}\times \mathbb{R}^3= \mathbf{y}=\mathbf{h}(v), \quad |v|<\delta\right\}; \quad \mathbf{h}(0)=0, D \mathbf{h}(0)= \mathbf{0},
\end{equation}
for $\delta$ sufficently small (cf. reference \cite{perko}, p 55). The restriction of the system  \eqref{eq.47}-\eqref{eq.48}  to the center manifold is
\begin{equation}
    \frac{d v}{d \tau}= C v + f(v, \mathbf{h}(v)). \label{A.51}
\end{equation}
According to Theorem 3.2.2 in Ref. \cite{center}, if the origin of \eqref{A.51} is stable, ~resp. unstable, then
the origin of  \eqref{eq.47}-\eqref{eq.48} is also stable, ~resp. unstable. Therefore, we have to find the local center
manifold, i.e., the problem reduces to the computation of $\mathbf{h}(v)$.

Substituting  $\mathbf{y}= \mathbf{h}(x)$ in \eqref{eq.48} and using the chain rule, $\frac{d \mathbf{y}}{d \tau}= D \mathbf{h}(v) \frac{d v}{d \tau}$, on can show that the function  $\mathbf{h}(x)$ that defines the local center manifold satisfies
\begin{equation}
    D \mathbf{h}(v)\left[ C v + f(v, \mathbf{h}(v))\right]- P \mathbf{h}(v)- \mathbf{g}(v, \mathbf{h}(v)) = \mathbf{0}.
\end{equation}
This condition allows for an approximation of $\mathbf{h}(x)$ by a Taylor series at $v=0$. Since $\mathbf{h}(0)=0, D \mathbf{h}(0)= \mathbf{0}$, then $\mathbf{h}(x)$ commences with quadratic terms.
We substitute
\begin{equation}
    \mathbf{h}(v)= \left(\begin{array}{c}
         h_1(v)  \\
         h_2(v) \\
         h_3(v)
    \end{array}\right)= \left(\begin{array}{c}
        a_1 v^2 + a_2 v^3 + \ldots + a_{N-1} v^N  + \mathcal{O}(v^{N+1})\\
        b_1 v^2 + b_2 v^3 + \ldots + b_{N-1} v^N  + \mathcal{O}(v^{N+1})\\
        c_1 v^2 + c_2 v^3 + \ldots + c_{N-1} v^N  + \mathcal{O}(v^{N+1})
    \end{array}\right)
\end{equation}
and we find $\mathbf{h}(v)=\mathbf{0}$ at any given order $N$ in the Taylor expansion.

Therefore, applying this procedure to \eqref{eq.46}, \eqref{eq.43}, \eqref{eq.44}, \eqref{eq.45}, we obtain that the dynamics on the center manifold of $P_3$ is governed by
\begin{align}\label{A.7}
 \frac{d v}{d\tau}=    \frac{(\lambda -4) v^2}{\bar{\omega}_{BD}}.
\end{align}
It is obvious that the origin $v = 0$ of \eqref{A.7} is asymptotically unstable (saddle
point). According to Theorem 3.2.2 in Ref. \cite{center}, the origin of previous is also  unstable (saddle
point).

\begin{figure}
    \centering
    \includegraphics[width=0.9\textwidth]{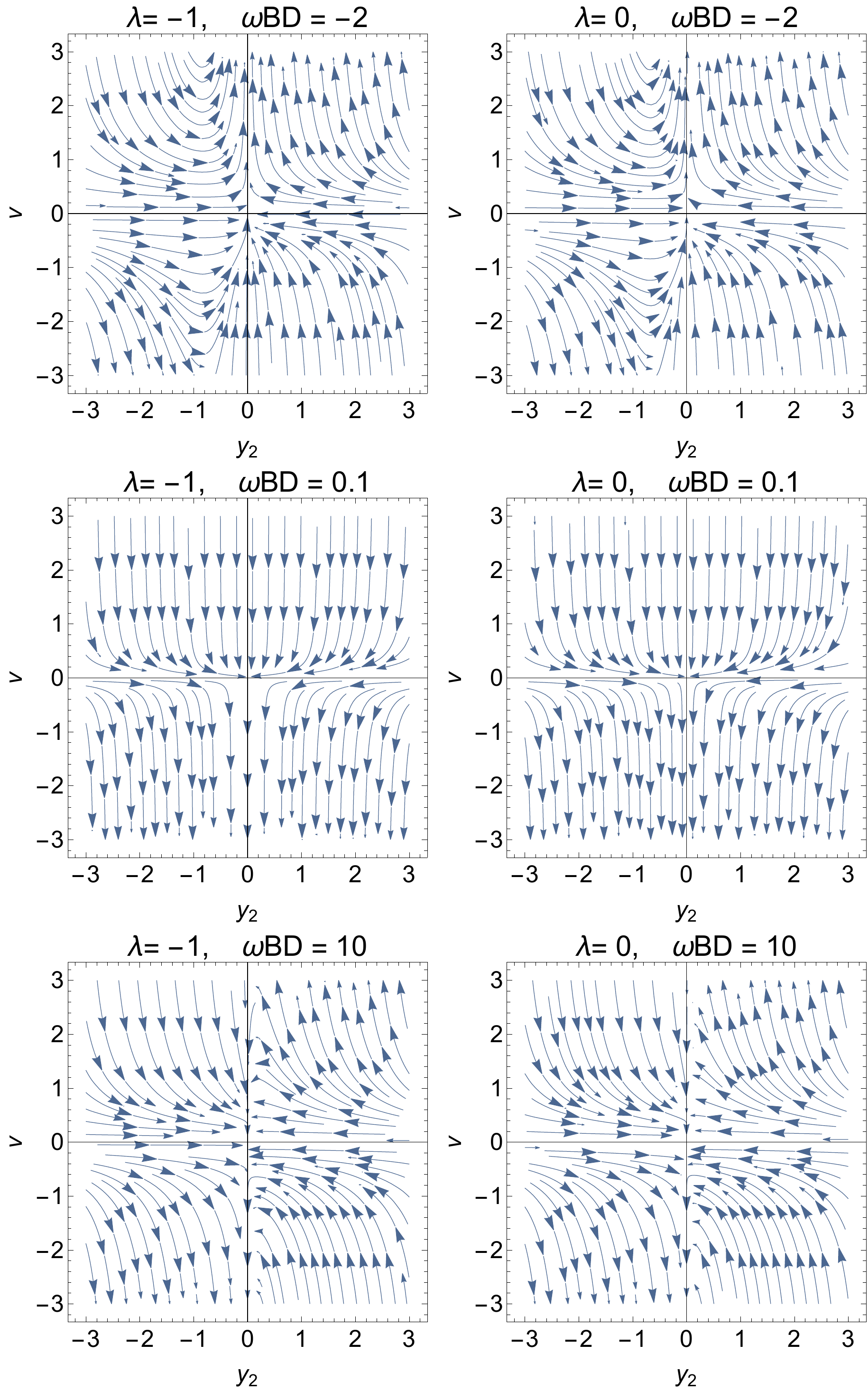}
    \caption{Phase plot of the dynamical system \eqref{system1}, where it is shown that the origin is unstable (saddle point).}
    \label{fig:1}
\end{figure}

For example, in the invariant set, $y_1=y_3=0$, the dynamics is given by
\begin{equation}
\label{system1}
    y_2'= -3 y_2, \quad v'=\frac{v^2 (\lambda +3 y_2\overline{\omega}_{BD}-4)}{\overline{\omega}_{BD}}.
\end{equation}
In figure \ref{fig:1} a phase plot of the dynamical system \eqref{system1} is presented, in which it is shown that the origin is unstable (saddle point).

\subsubsection{Case $\lambda=4$}

Consider now the case $\lambda=4$. The specific model can describe a
cosmological history with at least two acceleration phases, a matter epoch and
a radiation era. That is an interesting result as it can describe the important
eras of cosmological evolution. Note that $\lambda=4$ corresponds to
potential $V\left(  \psi\right)  =V_{0}\psi^{4},$ and for the original scalar
field $\phi$, the potential function $V\left(  \phi\right)  =V_{0}\phi^{2}$.

Now, for $\lambda=4$, the similarity matrix is given by
\begin{align}
    s= \left(
\begin{array}{cccc}
 0 & 0 & -\frac{1}{2} & 0 \\
 -\frac{3  \bar{\omega}_{BD}}{2} & 0 & 1 & 0 \\
 1 & 1 & 0 & 0 \\
 0 & 0 & 0 & 1 \\
\end{array}
\right).
\end{align}
Introduce the linear transformation
\begin{equation}
 \left(\begin{array}{c}
      v\\
      y_1 \\
      y_2 \\
      y_3
 \end{array}\right)  = \left(
\begin{array}{c}
\mu   \\
 -\frac{2 (2 x +y +1)}{3  \bar{\omega}_{BD}} \\
 \frac{2 (2 x +y +1)}{3  \bar{\omega}_{BD}}+z  \\
 -2 x
\end{array}
\right), \label{linear2}
\end{equation}
we obtain the system
\begin{align}
&\frac{d v}{d\tau}=  3 y_1 v^2+3    y_2 v^2-\frac{y_3 v}{2}, \label{eq.58}\\
 &\frac{d y_1}{d\tau}= -6 y_1+   y_1^2 (-6 y_3 v-18 v-9  \bar{\omega}_{BD}) \nonumber \\
 & +y_1 \left(-6 y_2 y_3 v+y_3^2 \left(v \left(\frac{2}{\bar{\omega}_{BD}}+\frac{1}{4}\right)+\frac{1}{2}\right)+y_3 \left(\frac{6 v}{ \bar{\omega}_{BD}}+6\right)\right)+\frac{2 y_2 y_3^2 v}{\bar{\omega}_{BD}} \nonumber \\
 & +y_3^3 \left(-\frac{v}{12  \bar{\omega}_{BD}}-\frac{1}{6  \bar{\omega}_{BD}}\right), \label{eq.59}
\\
    &\frac{d y_2}{d\tau}= -3 y_2 + y_1^2 \left(3 y_3 v+9 v+\frac{9\bar{\omega}_{BD}}{2}\right) \nonumber \\
    & +y_1 \Bigg(y_2 \left(-9 v-\frac{9  \bar{\omega}_{BD}}{2}\right)+y_3^2 \left(v \left(-\frac{2}{\bar{\omega}_{BD}}-\frac{1}{8}\right)-\frac{1}{4}\right) \nonumber \\
    & +y_3 \left(v \left(\frac{1}{2}-\frac{12}{ \bar{\omega}_{BD}}\right)-\frac{7}{2}\right)+v
   \left(\frac{3}{2}-\frac{18}{ \bar{\omega}_{BD}}\right)\Bigg)-3 y_2^2 y_3 v \nonumber \\
   & +y_2 \left(y_3^2 \left(v \left(\frac{1}{8}-\frac{2}{\bar{\omega}_{BD}}\right)+\frac{1}{4}\right)+y_3 \left(v \left(\frac{1}{2}-\frac{6}{ \bar{\omega}_{BD}}\right)+\frac{5}{2}\right)\right) \nonumber \\
   & +y_3^3 \left(\frac{v}{12
    \bar{\omega}_{BD}}+\frac{1}{6  \bar{\omega}_{BD}}\right)+y_3^2 \left(v \left(\frac{1}{4  \bar{\omega}_{BD}}-\frac{1}{48}\right)+\frac{1}{2 \bar{\omega}_{BD}}\right), \label{eq.60}
\\
    &\frac{d y_3}{d\tau}= -3 y_3 + y_1 \left(-3 y_3^2 v+y_3 \left(-3 v-\frac{9  \bar{\omega}_{BD}}{2}\right)+18 v\right)+y_2 \left(6 y_3 v-3
   y_3^2 v\right) \nonumber \\
   & +y_3^3 \left(\frac{v}{8}+\frac{1}{4}\right)+y_3^2 \left(\frac{5}{2}-\frac{v}{4}\right). \label{eq.61}
\end{align}
Applying the procedure to find the center manifold to \eqref{eq.58}, \eqref{eq.59}, \eqref{eq.60}, \eqref{eq.61}, we obtain that the dynamics on the center manifold of $P_3$ is governed by
\begin{align}\label{A.7}
 \frac{d v}{d\tau}=   0.
\end{align}
Therefore, we  rely on numerical inspection.
For example, in the invariant set $y_1=y_3=0$ the dynamics is given by
\begin{equation}
\label{system2}
    y_2'= -3 y_2, \quad v'=3v^2 y_2.
\end{equation}
    \begin{figure}
        \centering
        \includegraphics[width=0.7\textwidth]{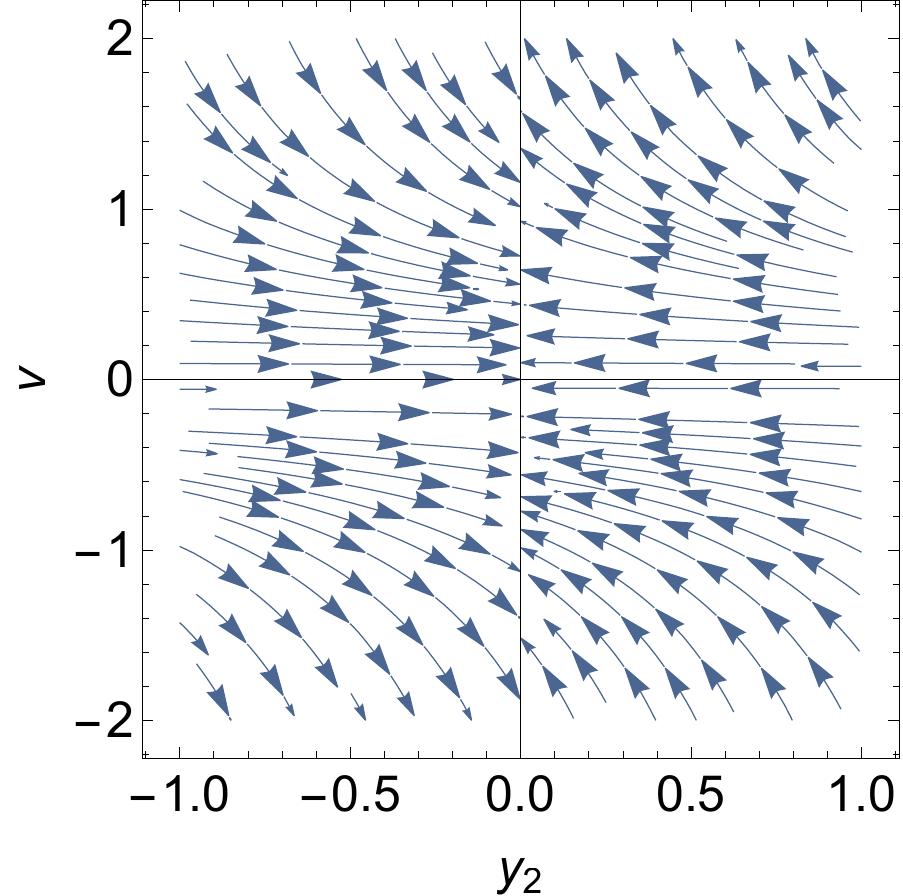}
        \caption{Phase plot of the dynamical system \eqref{system2}, wherein it is shown that the origin is unstable (saddle point). }
        \label{fig:2}
    \end{figure}
In figure \ref{fig:2} a phase plot of the dynamical system \eqref{system2}  is presented, where it is shown that the origin is unstable (saddle point).

\section{Nonzero spatial curvature}
\label{sec5}

We now assume the presence of nonzero spatial curvature, that is, the
background space is described by the line element%
\begin{equation}
ds^{2}=-dt^{2}+a^{2}\left(  t\right)  \left(  \frac{dr^{2}}{1-Kr^{2}}%
+r^{2}\left(  d\theta^{2}+\sin^{2}\theta d\phi^{2}\right)  \right).
\end{equation}
The parameter $K$ denotes the curvature for the three-dimensional hypersurface.
For a closed FLRW universe $K=+1$, while for an open FLRW universe $K=-1$.

For a nonzero $K$ the cosmological field equations are modified as follows%

\begin{equation}
6H^{2}+12H\left(  \frac{\dot{\psi}}{\psi}\right)  +\frac{\bar{\omega}_{BD}}%
{2}\left(  \dot{\psi}^{2}+2\beta\hbar^{2}\left(  \frac{\dot{\psi}}{\psi
}\right)  \left(  \frac{\dot{\zeta}}{\psi}\right)  -\frac{\zeta^{2}}{\psi^{2}%
}\right)  +\frac{V\left(  \psi\right)  }{\psi^{2}}-6K\frac{\psi^{2}}{a^{2}}=0,
\end{equation}%
\begin{equation}
2\dot{H}+3H^{2}+4\left(  \frac{\dot{\psi}}{\psi}\right)  H+\left(  \frac
{\bar{\omega}_{BD}}{4}-2\right)  \left(  \frac{\dot{\psi}}{\psi}\right)
^{2}+2\left(  \frac{\ddot{\psi}}{\psi}\right)  +\frac{V\left(  \psi\right)
}{2\psi^{2}}+\beta\frac{\bar{\omega}_{BD}}{4\psi^{2}}\left(  \zeta^{2}%
-2\dot{\zeta}\dot{\psi}\right)  -2K\frac{\psi^{2}}{a^{2}}=0,
\end{equation}%
\begin{equation}
\ddot{\psi}+3H\dot{\psi}+\zeta=0,
\end{equation}%
\begin{equation}
\bar{\omega}_{BD}\left(  \beta\ddot{\zeta}+\ddot{\psi}\right)  +3\bar{\omega
}_{BD}H\left(  \beta\dot{\zeta}+\dot{\psi}\right)  +V_{,\psi}+12\psi\left(
\dot{H}+2H^{2}-Ka^{-2}\right)  =0.
\end{equation}

In order to proceed with the analysis of the dynamical evolution and the
investigation of the stationary points for the latter system we define the new
set of variables \cite{kkd1}
\begin{equation}
X=\frac{\dot{\psi}}{\psi\sqrt{H^{2}+\left\vert K\right\vert a^{-2}}}%
~,~Y=\frac{V\left(  \psi\right)  }{6\psi^{2}\left(  H^{2}+\left\vert
K\right\vert a^{-2}\right)  }~,~Z=\beta\hbar^{2}\frac{\dot{\zeta}}{6\psi
^{2}\sqrt{H^{2}+\left\vert K\right\vert a^{-2}}}~~,
\end{equation}%
\begin{equation}
W=\beta\hbar^{2}\frac{\zeta^{2}}{12\psi^{2}\sqrt{H^{2}+\left\vert K\right\vert
a^{-2}}}~,~\eta=\frac{H}{\sqrt{H^{2}+\left\vert K\right\vert a^{-2}}%
}~,~\lambda=\psi\frac{V_{,\psi}\left(  \psi\right)  }{V\left(  \psi\right)
}~,
\end{equation}%
\[
\mu=-\frac{1}{\beta}\frac{\psi}{\zeta}~,~\Gamma\left(  \lambda\right)
=\frac{V_{,\psi\psi}\left(  \psi\right)  V\left(  \psi\right)  }{\left(
V_{,\psi}\left(  \psi\right)  \right)  ^{2}}. 
\]

In the following we consider the two cases, $K=+1$ and $K=-1$.

\subsection{Closed Universe}

For a positive curvature term, i.e. $K=+1$, and for the new independent
variable $d\sigma=\sqrt{H^{2}+\left\vert K\right\vert a^{-2}}dt$, the field
equations in the dimensionless variables are
\begin{align}
8\frac{dX}{d\sigma}  &  =48\mu W-4\eta X\left(  2\eta^{2}+3W(4\mu
+\text{$\bar{\omega}_{BD}$})+3Y-5\right) \nonumber\\
&  +(\text{$\bar{\omega}_{BD}$}-8)\eta X^{3}-4X^{2}\left(  2\eta
^{2}-3\text{$\bar{\omega}_{BD}$}\eta z+2\right),
\\
\frac{4}{Y}\frac{dY}{d\sigma}  &  =-4\eta\left(  2\eta^{2}+3W(4\mu
+\text{$\bar{\omega}_{BD}$})+3Y+1\right) \nonumber\\
&  +(\text{$\bar{\omega}_{BD}$}-8)\eta X^{2}+4X\left(  \lambda-2\eta
^{2}+3\text{$\bar{\omega}_{BD}$}\eta Z-2\right),
\\
-\text{$\bar{\omega}_{BD}$}\frac{dZ}{d\sigma}  &  =-4W(-2(\text{$\bar{\omega
}_{BD}$}-12)\mu-3\text{$\bar{\omega}_{BD}$}\eta Z(4\mu+\text{$\bar{\omega
}_{BD}$})+6\text{$\bar{\omega}_{BD}$})\nonumber\\
&  +4X\left(  -4\eta-3\text{$\bar{\omega}_{BD}$}^{2}\eta Z^{2}+2\text{$\bar
{\omega}_{BD}$}\left(  \eta^{2}+4\right)  Z\right) \nonumber\\
&  +(\text{$\bar{\omega}_{BD}$}-8)X^{2}(2-\text{$\bar{\omega}_{BD}$}\eta
Z)+8\left(  2\eta^{2}+(\lambda-3)Y-1\right) \nonumber\\
&  +4\text{$\bar{\omega}_{BD}$}\eta Z\left(  2\eta^{2}+3Y-5\right),
\\
\frac{4}{W}\frac{dW}{d\sigma}  &  =4\left(  -\eta\left(  2\eta^{2}%
+3W(4\mu+\text{$\bar{\omega}_{BD}$})+3Y+1\right)  -6\mu Z\right) \nonumber\\
&  +(\text{$\bar{\omega}_{BD}$}-8)\eta X^{2}-4X\left(  2\eta^{2}%
-3\text{$\bar{\omega}_{BD}$}\eta Z+2\right),
\end{align}%
\begin{equation}
\frac{8}{\left(  \eta^{2}-1\right)  }\frac{d\eta}{d\sigma}=\left(  8\eta
^{2}+12W(4\mu+\text{$\bar{\omega}_{BD}$})-X(-8\eta+(\text{$\bar{\omega}_{BD}$%
}-8)X+12\text{$\bar{\omega}_{BD}$}Z)+12y-4\right),
\end{equation}%
\begin{equation}
\frac{d\lambda}{d\sigma}=\lambda x\left(  1-\lambda\left(  1-\Gamma\left(
\lambda\right)  \right)  \right),
\end{equation}%
\begin{equation}
\frac{d\mu}{d\sigma}=\mu\left(  x+3z\mu\right),
\end{equation}
with constraint%
\begin{equation}
12\left(  Y+2\eta^{2}-1\right)  -24X\eta-12\bar{\omega}_{BD}W+\bar{\omega
}_{BD}x\left(  X+12Z\right)  =0. \label{cc.01}%
\end{equation}

As before for the scalar field potential we consider the power-law potential
function in which $\lambda=const$.

By using the algebraic equation (\ref{cc.01}) the stationary points for the
positive curvature are derived to be of the form $\mathbf{A}=\left(  X\left(
A\right),Y\left(  A\right),Z\left(  A\right),\mu\left(  A\right)
,\eta\left(  A\right)  \right)  $. Indeed the stationary points are%
\[
A_{1}^{\pm}=\left(  \pm1,0,Z_{1},\pm1\right),~A_{2}^{\pm}=\left(  \pm
\frac{6}{\lambda},\frac{2}{\lambda^{2}}\left(  \lambda-6\right),Z_{2}%
,\pm1\right),
\]%
\[
A_{3}^{\pm}=\left(  0,-1,\frac{\lambda-4}{3\bar{\omega}_{BD}},\pm1\right)
~,~A_{4}^{\pm}=\left(  \pm1,0,\pm\left(  \frac{1}{12}-\frac{1}{\bar{\omega
}_{BD}}\right),\frac{4\bar{\omega}_{BD}}{\bar{\omega}_{BD}-4},\pm1\right),
\]%
\[
A_{5}^{\pm}=\left(  \pm\frac{6}{\lambda},\frac{2}{\lambda^{2}}\left(
\lambda-6\right), \mp\frac{\lambda\left(  \lambda-10\right)  +3\left(
\bar{\omega}_{BD}-4\right)  }{6\bar{\omega}_{BD}\lambda},\frac{12\bar{\omega
}_{BD}}{\lambda\left(  \lambda-10\right)  +3\left(  \bar{\omega}%
_{BD}-4\right)  },\pm1\right),
\]%
\[
A_{6}^{\pm}=\left(  X_{6},0,-\frac{2}{\bar{\omega}_{BD}}\left(  \mp2+\frac
{1}{X_{6}}\right)  -\frac{X_{6}}{6},\frac{2\bar{\omega}_{BD}X_{6}^{2}%
}{12+X_{6}\left(  \mp24+\bar{\omega}_{BD}X_{6}\right)  }\pm1\right),
\]%
\[
A_{7}^{\pm}=\left(  0,0,0,\mu,\pm\frac{\sqrt{2}}{2}\right),~A_{8}^{\pm
}=\left(  \pm\sqrt{2},0,Z_{9},0,\pm\frac{\sqrt{2}}{2}\right),
\]%
\[
A_{9}^{\pm}=\left(  \pm\sqrt{2},0,\frac{\bar{\omega}_{BD}-12}{6\sqrt{2}%
\bar{\omega}_{BD}},\frac{4\bar{\omega}_{BD}}{\bar{\omega}_{BD}-12},\pm
\frac{\sqrt{2}}{2}\right).
\]
For each stationary point we calculate the equation of state parameter
$w_{tot}=-1-\frac{2}{3}\frac{\dot{H}}{H}$ and the deceleration parameter
$q=\frac{1}{2}\left(  1+3w_{tot}\right)  \left(  1+\frac{K}{a^{2}H^{2}%
}\right)  $.

Stationary points, $A_{1}^{\pm}$,~$A_{2}^{\pm}$, $A_{3}^{\pm}$, $A_{4}^{\pm}$
and $A_{5}^{\pm}$, describe spatially flat asymptotic solutions as described by
the stationary points $P_{1}$,~$P_{2}$, $P_{3}$, $P_{4}$ and $P_{5}$,
respectively. For the points $A_{6}^{\pm}$ we derive $w_{tot}\left(
A_{6}^{\pm}\right)  =\frac{1}{3}\left(  3+2X_{6}\left(  X_{6}\mp2\right)
\right)  $ and $q\left(  A_{6}^{\pm}\right)  =2\left(  1\mp X_{6}\right)  $.
Consequently, $A_{6}^{\pm}$ describe asymptotic solutions with curvature. For
the point $A_{6}^{+}$ the scale factor is exponential when $X_{6}=1\pm
\frac{\sqrt{10}}{2}$.

\begin{table*}[h] \centering
\caption{Stationary points for the modified Brans-Dicke cosmological model with positive curvature.}%
\begin{tabular}
[c]{cccc}\hline\hline
\textbf{Point} & $\mathbf{w}_{tot}\left(  \mathbf{A}\right)  $ &
\textbf{Curvature} & \textbf{Stability}\\\hline
$A_{1}^{\pm}$ & $\frac{1}{3}$ & Flat & Saddle\\
$A_{2}^{\pm}$ & $1-\frac{4}{\lambda}$ & Flat & Saddle\\
$A_{3}^{\pm}$ & $-1$ & Flat & Saddle\\
$A_{4}^{\pm}$ & $\frac{1}{3}$ & Flat & Saddle\\
$A_{5}^{\pm}$ & $1-\frac{4}{\lambda}$ & Flat & Saddle\\
$A_{6}^{\pm}$ & $\frac{1}{3}\left(  3+2X_{6}\left(  X_{6}\mp2\right)  \right)
$ & $K>0$ & Saddle\\
$A_{7}^{\pm}$ & $-\frac{1}{3}$ & $K>0$ & Saddle\\
$A_{8}^{\pm}$ & $-\frac{1}{3}$ & $K>0$ & Saddle\\
$A_{9}^{\pm}$ & $-\frac{1}{3}$ & $K>0$ & $A_{9}^{+}$ attractor for $\lambda
<3$\\\hline\hline
\end{tabular}
\label{tab2}%
\end{table*}%

The asymptotic solutions at the stationary points
$A_{7}^{\pm}$, $~A_{8}^{\pm}$ and $~A_{9}^{\pm}$ are that with $w_{tot}%
=-\frac{1}{3}$ and $q=0$. Hence, at the points the asymptotic solutions
describe spacetimes with nonzero spatial curvature.

We calculate the eigenvalues for the linearized system in order to study the
stability properties for the stationary points.

The linearized system around the points $A_{1}^{\pm}$ gives the eigenvalues
$e_{2}\left(  A_{1}^{\pm}\right)  =0$~, $e_{2}\left(  A_{1}^{\pm}\right)
=\pm1$~, $e_{3}\left(  A_{1}^{\pm}\right)  =\mp1~$,~$e_{4}\left(  A_{1}^{\pm
}\right)  =\pm2$~\ and $e_{5}\left(  A_{1}^{\pm}\right)  =\pm\left(
\lambda-6\right)  .~$Therefore, the solutions at the points $A_{1}^{\pm}$ are
always unstable and $A_{1}^{\pm}$ are saddle points.

The eigenvalues of the stationary points $A_{2}^{\pm}$ are derived to be
$e_{1}\left(  A_{2}^{\pm}\right)  =0$,$~e_{2}\left(  A_{2}^{\pm}\right)
=\pm\frac{6}{\lambda}$,$~e_{3}\left(  A_{2}^{\pm}\right)  =\pm\left(
4-\frac{12}{\lambda}\right)  ~$, $e_{4}\left(  A_{2}^{\pm}\right)  =\mp
\frac{3}{\lambda}\left(  1+\sqrt{25-4\lambda}\right)  $ and $e_{5}\left(
A_{2}^{\pm}\right)  =\frac{3}{\lambda}\left(  1+\sqrt{25-4\lambda}\right)  $.
Consequently, $A_{2}^{\pm}$ are saddle points.

For points $A_{3}^{\pm}$ the eigenvalues are $e_{1}\left(  A_{3}^{\pm}\right)
=0$~, $e_{2}\left(  A_{3}^{\pm}\right)  =\pm3$, $e_{3}\left(  A_{3}^{\pm
}\right)  =\pm3$, $e_{4}\left(  A_{3}^{\pm}\right)  =\mp2$, $e_{5}\left(
A_{3}^{\pm}\right)  =\pm6$.  Hence points $A_{3}^{\pm}$ are saddle points.

Similarly for the \ points $~A_{4}^{\pm}$ we derive $e_{1}\left(  A_{4}^{\pm
}\right)  =\mp1$, $~e_{2}\left(  A_{4}^{\pm}\right)  =\mp1$, $e_{3}\left(
A_{4}^{\pm}\right)  =\pm2\,$\, $e_{4}\left(  A_{4}^{\pm}\right)  =\mp2$,
$e_{5}\left(  A_{4}^{\pm}\right)  =\pm\left(  \lambda-6\right)  ~$, that is,
$A_{4}^{\pm}$ are saddle points.

The linearized system around points $A_{5}^{\pm}$ are~$e_{1}\left(  A_{5}%
^{\pm}\right)  =\mp\frac{6}{\lambda}$~, $e_{2}\left(  A_{5}^{\pm}\right)
=\pm\left(  4-\frac{12}{\lambda}\right)  $, $e_{3}\left(  A_{5}^{\pm}\right)
=\mp\frac{12}{\lambda}$ and $e_{4}\left(  A_{5}^{\pm}\right)  =\mp\frac
{3}{\lambda}\left(  1+\sqrt{\left(  25-4\lambda\right)  }\right)
~,~e_{5}\left(  A_{5}^{\pm}\right)  =\mp\frac{3}{\lambda}\left(
1-\sqrt{\left(  25-4\lambda\right)  }\right)  $, that is, the solutions at
points $A_{5}^{\pm}$ are always unstable. Points $A_{5}^{\pm}$ are always
saddle points.

The eigenvalues of points $A_{6}^{\pm}$ are derived to be $e_{1}\left(  A_{6}^{\pm
}\right)  =0$, $e_{2}\left(  A_{6}^{\pm}\right)  =-4\left(  X_{6}\mp1\right)
$,~$e_{3}\left(  A_{6}^{\pm}\right)  =\left(  2+\lambda\right)  X_{6}\mp6$,
$e_{4}\left(  A_{6}^{\pm}\right)  =X_{6}\sqrt{\frac{X_{6}\left(  \pm48+\left(
\bar{\omega}_{BD}-36\right)  X_{6}\right)  -24}{X_{6}\left(  \bar{\omega}%
_{BD}\mp24\right)  +12}}$, $e_{5}\left(  A_{6}^{\pm}\right)  =-X_{6}%
\sqrt{\frac{X_{6}\left(  \pm48+\left(  \bar{\omega}_{BD}-36\right)
X_{6}\right)  -24}{X_{6}\left(  \bar{\omega}_{BD}\mp24\right)  +12}}$. Thus,
the stationary points $A_{6}^{\pm}$ are saddle points.

For the points $A_{7}^{\pm}$ two of eigenvalues are $e_{1}\left(  A_{7}^{\pm
}\right)  =\sqrt{2}$~,~$e_{2}\left(  A_{7}^{\pm}\right)  =-\sqrt{2}$, from
where we infer that $A_{7}^{\pm}$ are always saddle points.

The five eigenvalues\ for the linearized system around points $A_{8}^{\pm}%
~$are $e_{1}\left(  A_{8}^{\pm}\right)  =0$, $e_{2}\left(  A_{8}^{\pm
}\right)  =\sqrt{2}~$, $e_{1}\left(  A_{8}^{\pm}\right)  =-\sqrt{2}$,
$e_{2}\left(  A_{8}^{\pm}\right)  =\mp2\sqrt{2}\,$, $e_{1}\left(  A_{8}^{\pm
}\right)  =\pm\sqrt{2}\left(  \lambda-3\right)$, which means that points
\ $A_{8}^{\pm}$ are saddle points.

The eigenvalues of the linearized system around~$A_{9}^{\pm}$ are
$e_{1}\left(  A_{9}^{\pm}\right)  =\mp\sqrt{2}$,$~e_{2}\left(  A_{9}^{\pm
}\right)  =\mp\sqrt{2}$, $e_{3}\left(  A_{9}^{\pm}\right)  =\mp2\sqrt{2}$,
$e_{4}\left(  A_{9}^{\pm}\right)  =\mp2\sqrt{2}$, $e_{5}\left(  A_{9}^{\pm
}\right)  =\mp\sqrt{2}\left(  3-\lambda\right)  $, that is, point $A_{9}^{+}$
is an attractor for $\lambda<3$, while point $A_{9}^{+}$ is a source for
$\lambda<3$, otherwise is a saddle point.\ The results are summarized in Table
\ref{tab2}.%

\subsection{Open Universe}

For a negative curvature FLRW background space the field equations in the
dimensionless variables are%
\begin{align}
\frac{dX}{d\sigma}  &  =6\mu W-\frac{3}{2}\eta X(W(4\mu+\text{$\bar{\omega
}_{BD}$})+Y-1)\nonumber\\
&  +\frac{1}{8}(\text{$\bar{\omega}_{BD}$}-8)\eta X^{3}-\frac{1}{2}%
X^{2}\left(  2\eta^{2}-3\text{$\bar{\omega}_{BD}$}\eta Z+2\right),
\\
\frac{4}{Y}\frac{dY}{d\sigma}  &  =-12\eta(W(4\mu+\text{$\bar{\omega}_{BD}$%
})+Y+1)\nonumber\\
&  +(\text{$\bar{\omega}_{BD}$}-8)\eta X^{2}+4X\left(  \lambda-2\eta
^{2}+3\text{$\bar{\omega}_{BD}$}\eta Z-2\right),
\\
-8\bar{\omega}_{BD}\frac{dZ}{d\sigma}  &  =-4W(-2(\text{$\bar{\omega}_{BD}$%
}-12)\mu-3\text{$\bar{\omega}_{BD}$}\eta Z(4\mu+\text{$\bar{\omega}_{BD}$%
})+6\text{$\bar{\omega}_{BD}$})\nonumber\\
&  +4X\left(  -4\eta-3\text{$\bar{\omega}_{BD}$}^{2}\eta Z^{2}+2\text{$\bar
{\omega}_{BD}$}\left(  \eta^{2}+4\right)  Z\right)  +(\text{$\bar{\omega}%
_{BD}$}-8)X^{2}(2-\text{$\bar{\omega}_{BD}$}\eta Z)\nonumber\\
&  +8(\lambda-3)Y+12\text{$\bar{\omega}_{BD}$}\eta(Y-1)Z+8,
\\
\frac{4}{W}\frac{dW}{d\sigma} & =12(-\eta(W(4\mu+\text{$\bar{\omega}_{BD}$%
})+Y+1)-2\mu Z)+(\text{$\bar{\omega}_{BD}$}-8)\eta X^{2}\nonumber \\
& -4X\left(  2\eta
^{2}-3\text{$\bar{\omega}_{BD}$}\eta Z+2\right),
\\
\frac{d\lambda}{d\sigma}& =\lambda x\left(  1-\lambda\left(  1-\Gamma\left(
\lambda\right)  \right)  \right),
\\
\frac{d\mu}{d\sigma} & =\mu\left(  x+3z\mu\right),
\end{align}
\begin{align}
\frac{d\eta}{d\sigma}  &  =\frac{1}{8}\left(  \eta^{2}-1\right)
(12W(4\mu+\text{$\bar{\omega}_{BD}$})\nonumber\\
&  -X(-8\eta+(\text{$\bar{\omega}_{BD}$}-8)x+12\text{$\bar{\omega}_{BD}$%
}Z)+12Y+4).
\end{align}

Furthermore, the algebraic constraint equation is%
\begin{equation}
12\left(  1+Y\right)  +\bar{\omega}_{BD}X\left(  X+12Z\right)  -24X\eta
-12\bar{\omega}_{BD}W=0.
\end{equation}

For the power-law potential for which $\lambda=const$, the stationary points in
the five-dimensional space $\left(  X,Y,Z,\mu,\eta\right)  $ are points
$A_{1}^{\pm}$, $A_{2}^{\pm}$,~$A_{3}^{\pm}$,~$A_{4}^{\pm}$, $A_{5}^{\pm}~$and
$A_{6}^{\pm}$. $\ $The physical properties of the points are the same as those
 for positive curvature, as also the stability properties. What is more
important to mention here is that the Milne solution \cite{sm1} is not
provided by the GUP modified Brans-Dicke model. That is not an unexpected
result. Milne solution is the vacuum solution of General Relativity, but
because of the nature of coupling of the scalar field with the gravitational
field a zero contribution of the scalar field in the field equations is not allowed.

\section{Conclusions}

\label{sec6}

In this piece of work we proposed a modified Brans-Dicke cosmological model
inspired by the minimum length uncertainty. Specifically, in the Brans-Dicke
Action Integral the kinetic part of the scalar field has been modified in
order that the equation of motion for the scalar field be given by the
quadratic GUP. New higher-order derivative terms for the scalar field have
been introduced in the gravitational Action Integral. With the use of a
Lagrange multiplier the higher-order derivative terms have been attributed to
a new scalar field with nonzero interaction terms with the Brans-Dicke field.

We calculated the cosmological field equations for the a homogeneous and
isotropic background space, described by the FLRW metric. In order to
investigate the effects of the higher-order terms provided by GUP in the
cosmological evolution we considered dimensionless variables in the context of
the $H$-normalization and we determined the stationary points and their
stability properties. We perform our analysis for the spatially flat FLRW
universe as also in the presence of a nonzero spatial curvature.

We compare the results for the modified Brans-Dicke theory with that of the
unmodified theory, where we observe that the new degrees of freedom given by
GUP change dramatically the dynamics and the provided asymptotic solutions for
the field equations. Specifically, for the case of a power-law potential
function we found that more than one acceleration phases are provided by the
theory, as also an asymptotic solution which describes the radiation era is
always present. The physical properties of the stationary points depend upon on the
exponent of the potential function and not on the value of the Brans-Dicke
parameter, in contrary to the unmodified model.

This analysis is based on a series of studies where we consider the
modification of the scalar field Lagrangian inspired by the GUP. In previous
studies we consider the quintessence model while here we assume the
Brans-Dicke theory. It both cases we found that the higher-order terms
provided by GUP affects the dynamics such that more acceleration asymptotic
solutions to be provided in the cosmological dynamics. In addition, the de
Sitter universe is provided as an asymptotic solutions for the both theories,
either for scalar field potentials where the de Sitter universe does not exist
for the unmodified scalar field theories.

\begin{acknowledgments}
This work is based on the research supported in part by the National Research
Foundation of South Africa (Grant Number 131604). Additionally, this research is funded by Vicerrector\'ia de Investigaci\'on y Desarrollo Tecnol\'ogico at Universidad Cat\'olica del Norte.
\end{acknowledgments}

\end{document}